\documentclass[12pt]{article}

\usepackage[english]{babel}
\usepackage{graphicx,rotating,epsfig}
\usepackage{a4p}

\def\GeV  {\ensuremath{\mathrm{ Ge\kern -0.1em V } }}
\def\GeVc2{\ensuremath{\mathrm{ Ge\kern -0.1em V }\kern -0.2em /c^2 }}
\newcommand{\MT}{\ensuremath{M_{\mathrm{top}}}}
\newcommand{\MW}{\ensuremath{M_{\mathrm{ W }}}}
\newcommand{\Pt}{\ensuremath{t}}
\newcommand{\Ptt}{\ensuremath{\Pt\bar\Pt}}

\begin{document}


\begin{center}
{\LARGE FERMI NATIONAL ACCELERATOR LABORATORY}
\end{center}

\begin{flushright}
       TEVEWWG/top 2005/01\\
       FNAL TM-2321-E\\
       CDF Note 7727 \\
       D\O\ Note 4864 \\
       hep-ex/0507006 \\
       {\bf 30th June 2005}
\end{flushright}

\vskip 1cm

\begin{center}
{\Huge \bf Combination of CDF and D\O\ Results \\[3mm]
                  on the Top-Quark Mass}
\vskip 1cm
{\Large
The CDF Collaboration, the D\O\ Collaboration, and \\[1mm]
the Tevatron Electroweak Working Group\footnote{
WWW access at {\tt http://tevewwg.fnal.gov}\\
The members of the TEVEWWG who contributed significantly to the
analysis described in this note are:
J.~F.~Arguin (arguin@fnal.gov)          
F.~Canelli (canelli@fnal.gov),          
R.~Demina (demina@fnal.gov),            
I.~Fleck (fleck@fnal.gov),              
D.~Glenzinski (douglasg@fnal.gov),      
E.~Halkiadakis (evah@fnal.gov),         
M.~W.~Gr\"unewald (mwg@fnal.gov),       
A.~Juste (juste@fnal.gov),              
T.~Maruyama (maruyama@fnal.gov),        
A.~Quadt (quadt@fnal.gov),              
E.~Thomson (thomsone@fnal.gov),         
C.~Tully (tully@fnal.gov),              
E.~W.~Varnes (varnes@fnal.gov).         
}
}

\vskip 1cm

{\bf Abstract}

\end{center}

{
  
  The results on the measurements of the top-quark mass, based on the
  data collected by the Tevatron experiments CDF and D\O\ at Fermilab
  during Run I from 1992 to 1996, and Run II since 2001 are
  summarized. The combination of the published Run I and preliminary
  Run II results, taking correlated uncertainties properly into
  account, is presented.  The resulting preliminary world average for
  the mass of the top quark is: $\MT=174.3 \pm 3.4~\GeVc2$, where the
  total error consists of a statistical part of $2.0~\GeVc2$ and a
  systematic part of $2.8~\GeVc2$.

}

\vfill



\section{Introduction}

The experiments CDF and D\O, taking data at the Tevatron
proton-antiproton collider located at the Fermi National Accelerator
Laboratory, have made several direct experimental measurements of the
pole mass, $\MT$, of the top quark $\Pt$.  The published measurements
~\cite{Mtop1-CDF-di-l-PRLa, Mtop1-CDF-di-l-PRLb,
  Mtop1-CDF-di-l-PRLb-E, Mtop1-D0-di-l-PRL, Mtop1-D0-di-l-PRD,
  Mtop1-CDF-l+j-PRL, Mtop1-CDF-l+j-PRD, Mtop1-D0-l+j-old-PRL,
  Mtop1-D0-l+j-old-PRD, Mtop1-D0-l+j-new, Mtop1-CDF-all-j-PRL,
  Mtop1-D0-all-j-PRL} are based on Run I data (1992-1996) while the
results from Run II are
preliminary~\cite{Mtop2-CDF-di-l,Mtop2-CDF-l+j,Mtop2-CDF-l+j-new,Mtop2-CDF-ll-new,Mtop2-D0-di-l,Mtop2-D0-l+j}.
They utilize all decay topologies\footnote{Decay channels with
  explicit identification of tau leptons in the final state are
  presently under study for cross section and branching ratio
  measurements. They are not yet used for measurements of the top
  quark mass.} arising in $\Ptt$ production given by the leptonic or
hadronic decay of the W boson occurring in top-quark decay: the
di-lepton channel (di-l)~\cite{Mtop1-CDF-di-l-PRLa,
  Mtop1-CDF-di-l-PRLb, Mtop1-CDF-di-l-PRLb-E, Mtop1-D0-di-l-PRL,
  Mtop1-D0-di-l-PRD,Mtop2-CDF-di-l,Mtop2-CDF-ll-new,Mtop2-D0-di-l},
the lepton+jets channel (l+j)~\cite{Mtop1-CDF-l+j-PRL,
  Mtop1-CDF-l+j-PRD, Mtop1-D0-l+j-old-PRL, Mtop1-D0-l+j-old-PRD,
  Mtop1-D0-l+j-new,Mtop2-CDF-l+j,Mtop2-CDF-l+j-new,Mtop2-D0-l+j}, and
the all-jets channel
(all-j)~\cite{Mtop1-CDF-all-j-PRL,Mtop1-D0-all-j-PRL}. The lepton+jets
channel yields the most precise determination of $\MT$.  The recently
presented preliminary measurement in this channel by the CDF
collaboration~\cite{Mtop2-CDF-l+j-new} is based on a large data set
with well controlled systematic uncertainties, yielding a top quark
mass precision better than the previous Run I world
average~\cite{Mtop1-tevew04}.

This note reports on the combination of a subset of those measurements. For
simplicity, we chose to combine the single best measurement in the lepton+jets
and the dilepton channel from each experiment, in particular the most recent
preliminary Run II measurements from
CDF~\cite{Mtop2-CDF-l+j-new,Mtop2-CDF-ll-new}, and the published Run I
measurements from D\O~\cite{Mtop1-D0-di-l-PRD,Mtop1-D0-l+j-new}. These are the
measurements with the largest weight and for which information on error
correlations is available. In the future, as our understanding of all the
error correlations matures and results become available, we plan to include
more measurements, from Run I and Run II, in the combination. 

The combination takes into account the statistical and systematic
uncertainties as well as the correlations between systematic
uncertainties, and replaces the previous
combination~\cite{Mtop1-tevew04}. The new CDF
measurement~\cite{Mtop2-CDF-l+j-new} in the lepton+jets channel is the
single most precise top-quark mass measurement and has the largest
weight in this new combination.

\section{Measurements}


The four measurements of $\MT$ to be combined are listed in
Table~\ref{tab:inputs}. The new CDF measurement in the lepton+jets
channel constrains the jet energy scale simultaneously from external
studies (calorimeter-track comparisons on $E/p$ from single isolated
tracks) as well from an in-situ calibration, based on the hadronic
$W\rightarrow q q^\prime$ invariant mass in the $t\bar{t}$ events.

For the combination procedure the CDF lepton+jets channel is split up
into two separate measurements with identical central value and fully
correlated statistical and systematic errors. Only the jet energy
scale uncertainty is uncorrelated. One measurement, (CDF-II\ 
l+j)$\rm_e$, is associated with an energy scale uncertainty of
$3.1\;\rm GeV$ from the external calibration, which is fully
correlated with the CDF measurement in the dilepton channel. The other
measurement, (CDF-II l+j)$\rm_i$, is associated with an energy scale
uncertainty of $4.2\;\rm GeV$, estimated to be the contribution from
the in-situ calibration, which is uncorrelated with any CDF or D\O\ 
measurement.  The combination of these two measurements yields
identical central value, statistical error and systematic error, and a
total jet energy scale uncertainty of $2.5\;\rm GeV$, as quoted
in~\cite{Mtop2-CDF-l+j-new} for the CDF measurement in the lepton+jets
channel.

Besides central values and statistical uncertainties, the systematic
errors arising from various sources are reported in
Table~\ref{tab:inputs}. In order of decreasing importance, the
systematic error sources are:
\begin{itemize}
\item Jet energy scale: The systematics for jet energy scale include
  the uncertainties on the absolute jet energy corrections,
  calorimeter stability, underlying event and relative jet energy
  corrections. Since the jet energy scale uncertainty is the largest
  uncertainty in all channels, dominating the overall precision of
  this combination, the various components of this uncertainty have
  been studied quantitatively and are grouped into contributions, 
  correlated or uncorrelated between the two experiments.
  
  {\bf iJES :} The component of the jet energy scale originating from
  in-situ calibration procedures, here using the $W \rightarrow
  qq^\prime$ invariant mass in the CDF-II\ l+j channel, is labeled
  iJES and treated uncorrelated with any other channel.
  
  {\bf bJES :} The component of the jet energy scale covering aspects
  of the $b$-jet energy scale, in particular fragmentation, color flow
  and semileptonic decay fractions, is labeled bJES and treated fully
  correlated with all channels of all experiments.
  
  {\bf cJES :} The correlated part of the remaining, external jet
  energy scale uncertainty (cJES) from the external calibration
  includes uncertainties from fragmentation and out-of-cone showering
  corrections and is correlated between all channels of all
  experiments.
  
  {\bf rJES :} The uncorrelated part of the remaining, external jet
  energy scale uncertainty (rJES) summarizes uncertainties mainly from
  the calorimeter response, relative response of different calorimeter
  sections, multiple interactions for CDF and contributions from the
  underlying event. It is treated as correlated between all channels
  in a given experiment, but not between experiments.
  
  Further studies, in particular on the breakdown of the various
  contributions to the jet energy scale uncertainties and their
  correlations are necessary to achieve better understanding and will
  be pursued in the future. The described procedure with the quoted
  numbers represent our current, preliminary understanding of the jet
  energy scale uncertainty and its correlation.

\item Model for signal (signal): The systematics for the signal model
  include initial and final state radiation effects, b-tagging bias,
  dependence upon parton distribution functions as well as variations
  in $\Lambda_{\mathrm{QCD}}$.

\item Model for background (BG): The background model includes
  estimates of the effect of varying the fragmentation scale from
  $Q^2=\MW^2$ to $Q^2=\langle p_t\rangle^2$ in VECBOS~\cite{VECBOS}
  simulations of W+jets production, the use of ISAJET~\cite{ISAJET}
  fragmentation instead of HERWIG~\cite{HERWIG5} fragmentation as well
  as the effect of varying the background fraction attributed to QCD
  multijet production with fake leptons and missing $E_T$.
  
\item Uranium noise and multiple interactions (UN/MI): This
  uncertainty includes uncertainties arising from uranium noise in the
  D\O\ calorimeter and from multiple interactions overlapping signal
  events. CDF includes the systematic uncertainty due to multiple
  interactions in the uncorrelated part of the JES contribution from
  external calibration.
  
\item Method for mass fitting (fit): This systematic uncertainty takes
  into account the finite sizes of Monte Carlo samples used for
  fitting, impact of jet permutations, and other fitting biases.  In
  the CDF lepton+jets analysis, the systematic uncertainty due to
  finite Monte Carlo statistics is included in the statistical
  uncertainty.
  
\item Monte Carlo generator (MC): The systematic uncertainty on the
  Monte Carlo generator provides an estimate of the sensitivity to the
  simulated physics model by comparing HERWIG to
  PYTHIA~\cite{PYTHIA4,PYTHIA5} or to ISAJET.  In the D\O\ analyses,
  the systematic uncertainty associated with the comparison of HERWIG
  to ISAJET is included in the signal model uncertainty.

\end{itemize}
For each measurement, the individual error contributions are combined
in quadrature.

\begin{table}[ht]
\begin{center}
\renewcommand{\arraystretch}{1.30}
\begin{tabular}{|l||rrr|rr|}
\hline
       &(CDF-II l+j)$\rm_i$ &(CDF-II l+j)$\rm_e$ & CDF-II di-l &  D\O\-I l+j & D\O\-I di-l \\
\hline
\hline
Result &  173.5 &   173.5 &    165.3 &      180.1 &     168.4 \\
\hline                       
\hline                       
Stat. &     2.7 &     2.7 &      6.3 &        3.6 &      12.3 \\
\hline                       
\hline                       
iJES  &     4.2 &     0.0 &      0.0 &        0.0 &       0.0 \\
bJES  &     0.0 &     0.6 &      0.8 &        0.6 &       0.8 \\
cJES  &     0.0 &     2.0 &      2.2 &        2.0 &       2.2 \\
rJES  &     0.0 &     2.3 &      1.4 &        2.6 &       0.5 \\

Signal&     1.1 &     1.1 &      1.5 &        1.1 &       1.8 \\
BG    &     1.2 &     1.2 &      1.6 &        1.0 &       1.1 \\
UN/MI &     0.0 &     0.0 &      0.0 &        1.3 &       1.3 \\
Fit   &     0.6 &     0.6 &      0.6 &        0.6 &       1.1 \\
MC    &     0.2 &     0.2 &      0.8 &        0.0 &       0.0 \\
\hline                       
Syst. &     4.5 &     3.5 &      3.6 &        3.9 &       3.6 \\
\hline                       
\hline                       
Total &     5.3 &     4.4 &      7.3 &        5.3 &      12.8 \\
\hline
\end{tabular}
\end{center}
\caption[Input measurements]{Summary of the five measurements of 
$\MT$ performed by CDF and D\O in two channels. All numbers are in 
$\GeVc2$. Note that the CDF measurement in the lepton+jets channel is 
devided into two measurements in order to treat the correlations of the 
jet energy scale uncertainties properly. For each measurement, the 
corresponding column lists experiment and channel, central value and 
contributions to the total error, namely statistical error and 
systematic errors arising from various sources defined in the text. 
Overall systematic errors and total errors are obtained by combining 
individual errors in quadrature.}
\label{tab:inputs}
\end{table}

\section{Combination}

In the combination, the error contributions arising from different
sources are uncorrelated between measurements.  The correlations of
error contributions arising from the same source are as follows:
\begin{itemize} 
\item uncorrelated: statistical error, fit error, iJES error;
\item 100\% correlated within each experiment : rJES error, UN/MI error;
\item 100\% correlated within each channel: BG error;
\item 100\% correlated between all measurements : bJES error, cJES error, signal error, MC error.
\end{itemize}
Note that the jet energy scale uncertainty from the in-situ
calibration (iJES) in the CDF lepton+jets measurement is treated as
uncorrelated with any other measurement. All other uncertainties are
treated fully correlated between the two CDF lepton+jets measurements.
The resulting matrix of global correlation coefficients is listed in
Table~\ref{tab:correl}.

\begin{table}[ht]
\begin{center}
\renewcommand{\arraystretch}{1.30}
\begin{tabular}{|l||rrr|rr|}
\hline
       &(CDF-II l+j)$\rm_i$ &(CDF-II l+j)$\rm_e$ & CDF-II di-l &  D\O\-I l+j & D\O\-I di-l \\
\hline
\hline
(CDF-II l+j)$\rm_i$   &    1.00 &           &          &           &         \\
(CDF-II l+j)$\rm_e$   &    0.43 &     1.00  &          &           &         \\
CDF-II di-l           &    0.05 &     0.30  &     1.00 &           &         \\
\hline                                                                       
D\O\ l+j              &    0.08 &     0.28  &     0.17 &      1.00 &         \\
D\O\ di-l             &    0.03 &     0.12  &     0.11 &      0.15 &    1.00 \\
\hline
\end{tabular}
\end{center}
\caption[Global correlations between input measurements]{Matrix of
global correlation coefficients between the five measurements of
$\MT$.}
\label{tab:correl}
\end{table}

The five measurements in the two channels are combined using a program
implementing a numerical $\chi^2$ minimization as well as the analytic
BLUE method~\cite{Lyons:1988, Valassi:2003}. The two methods used are
mathematically equivalent, and are also equivalent to the method used
in previous combinations~\cite{TM-2084}, and give identical results
for the combination. In addition, the BLUE method yields the
decomposition of the error on the average in terms of the error
categories specified for the input measurements~\cite{Valassi:2003}.

\section{Results}

The combined value for the top-quark mass is:
\begin{eqnarray}
\MT & = & 174.3 \pm 3.4~\GeVc2\,,
\end{eqnarray}
where the total error of $3.4~\GeVc2$ contains the following
components: a statistical error of $2.0~\GeVc2$; and systematic error
contributions of: total JES $2.2~\GeVc2$, signal $1.1~\GeVc2$,
background $1.0~\GeVc2$, UN/MI $0.4~\GeVc2$, fit $0.4~\GeVc2$, and MC
$0.2~\GeVc2$, for a total systematic error of $2.8~\GeVc2$.

\begin{figure}[t]
\begin{center}
\includegraphics[width=0.8\textwidth]{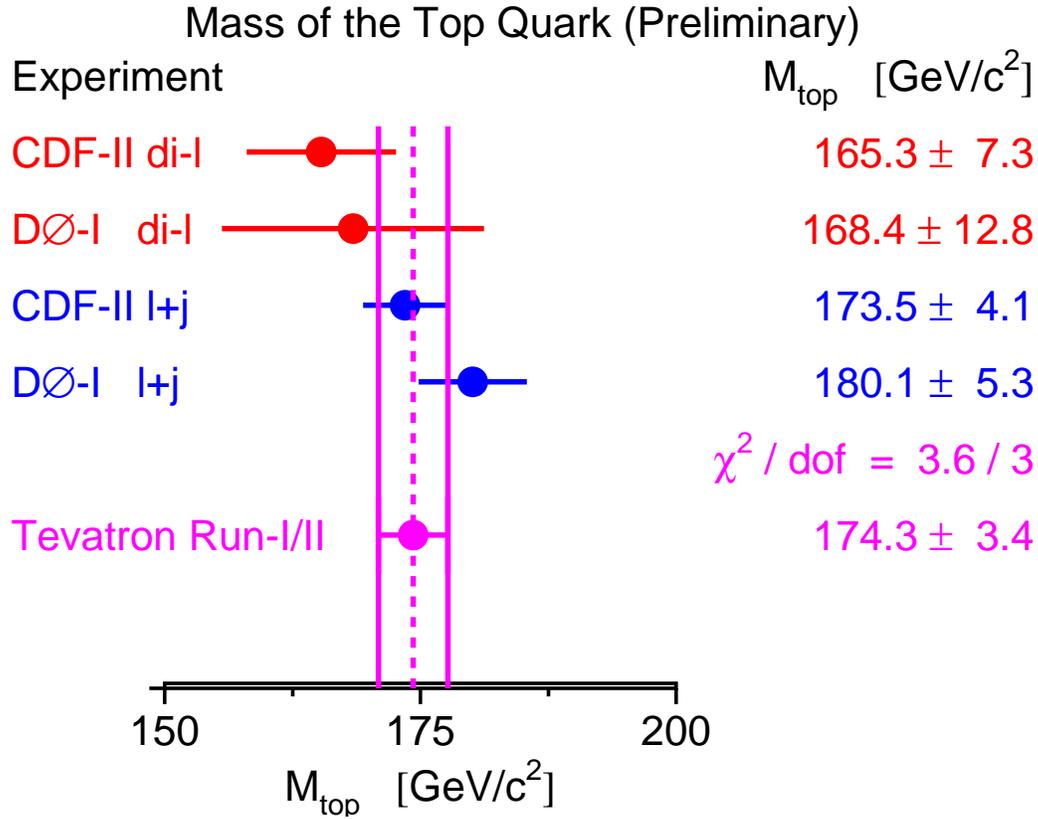}
\end{center}
\caption[Comparison of the measurements of the top-quark mass]
{Comparison of the measurements of the top-quark mass and their
average.}
\label{fig:mtop-bar-chart} 
\end{figure}

The $\chi^2$ of this average is 3.6 for 3 degrees of freedom,
corresponding to a probability of 47\%, showing that all measurements
are in good agreement with each other which can also be seen in
Figure~\ref{fig:mtop-bar-chart}.  The pull of each measurement with
respect to the average and the weight of each measurement in the
average are reported in Table~\ref{tab:stat}.


\begin{table}[t]
\begin{center}
\vskip 7mm
\renewcommand{\arraystretch}{1.30}
\begin{tabular}{|l||rrr|rr|}
\hline
       &(CDF-II l+j)$\rm_i$ &(CDF-II l+j)$\rm_e$ & CDF-II di-l &  D\O-I l+j & D\O-I di-l \\
\hline
\hline
Pull      & $-0.19$ &  $-0.27$ &   $-1.40$ &  $+1.44$ &   $-0.47$ \\
\hline
Weight    &  0.285  &   0.277  &    0.119  &   0.287  &    0.032  \\
\hline
\end{tabular}
\end{center}
\caption[Pull and weight of each measurement]{Pull and weight of each
  measurement in the average.}
\label{tab:stat} 
\end{table}

In addition, a combination of the five measurements with two physical
observables, the top quark mass in the lepton+jets decay channel,
$M_{top}^{l+j}$, and the top quark mass in the dilepton decay channel,
$M_{top}^{di-l}$, has been performed. The results of this combination,
obtained with a $\chi^2$ of 1.3 for 3 degrees of freedom and an
observed 28\% correlation between the two observables, are shown in
Table~\ref{tab:two_observables}.

\begin{table}[ht]
\begin{center}
\renewcommand{\arraystretch}{1.30}
\begin{tabular}{|l||r|r|}
\hline
       & $M_{top}^{l+j}$   &   $M_{top}^{di-l}$\\
\hline
\hline
Result &  175.7 &   166.1  \\
\hline                    
\hline                    
Stat. &     2.2 &     5.6  \\
\hline                    
\hline                    
total JES&  2.3 &     2.2  \\
Signal&     1.1 &     1.5  \\
BG    &     1.1 &     1.5  \\
UN/MI &     0.4 &     0.3  \\
Fit   &     0.5 &     0.5  \\
MC    &     0.2 &     0.6  \\
\hline                    
Syst. &     2.8 &     3.2  \\
\hline                       
\hline                       
Total &     3.6 &     6.4  \\
\hline
\end{tabular}
\end{center}
\caption[Input measurements]{Summary of the combination of the 5 measurements
by CDF and D\O\ in terms of two physical quantities, the top quark mass in the
lepton+jets and in the dilepton decay channel.}
\label{tab:two_observables}
\end{table}

The pull of each measurement with respect to the average of the two
physical observables and the weight of each measurement in the two
averages are reported in Table~\ref{tab:stat_two_observables}. Note
how the weights of the measurements of one channel add to zero in the
determination of the average for the other channel.  The two averages
are correlated (+28\%) through systematics.

\begin{table}[t]
\begin{center}
\vskip 7mm
\renewcommand{\arraystretch}{1.30}
\begin{tabular}{|l||rrr|rr|}
\hline
       &(CDF-II l+j)$\rm_i$ &(CDF-II l+j)$\rm_e$ & CDF-II di-l &  D\O-I l+j & D\O-I di-l \\
\hline
\hline
Pull(l+j) & $-0.57$ &  $-0.83$ &           &  $+1.11$ &           \\
Pull(di-l)&         &          &  $-0.23$  &          &  $0.21$   \\
\hline
Weight(l+j)&  0.288  &  0.377  &    0.0002 &   0.335  &  -0.0002  \\
Weight(di-l)& 0.268  & -0.283  &    0.788\,\,\, &   0.015  &   0.212\,\,\,  \\
\hline
\end{tabular}
\end{center}
\caption[Pull and weight of each measurement in the combination using two 
         physical observables]{Pull and weight of each
         measurement in the average.}
\label{tab:stat_two_observables} 
\end{table}

\section{Summary}

A preliminary combination of two published Run I measurements of $\MT$
from D\O\ and two preliminary Run II measurements from CDF is
presented. Taking into account statistical and systematic errors
including their correlations, the Tevatron and thus the world-average
result is: $\MT= 174.3 \pm 3.4~\GeVc2$. The mass of the top quark is now
known with an accuracy of 2.0\%.

\clearpage



  
\end{document}